\documentclass{article}
\usepackage{spconf,amsmath,graphicx}
\usepackage{url}
\usepackage{subfig}
\usepackage{xcolor}
\usepackage{cleveref}
\usepackage{booktabs, multirow} 
\usepackage{soul}
\usepackage{amsfonts}
\usepackage{cite}
\usepackage{caption}
\usepackage{subcaption}

\newcommand{\kaldi}{\texttt{Kaldi}}
\newcommand{\espnet}{\texttt{ESPnet}}
\newcommand{\espnetez}{\texttt{ESPnet-EZ}}

\title{ESPnet-EZ: Python-only ESPnet for Easy Fine-tuning and Integration}
\name{\parbox{\linewidth}{\centering Masao Someki*$^1$\thanks{* Equal contribution. Other authors are ordered by their surnames.}, Kwanghee Choi*$^1$, Siddhant Arora$^1$, William Chen$^1$, Samuele Cornell$^1$,\\Jionghao Han$^1$, Yifan Peng$^1$, Jiatong Shi$^1$, Vaibhav Srivastav$^2$, Shinji Watanabe$^1$}}
\address{$^1$Carnegie Mellon University, USA \,\,
$^2$Hugging Face, USA }

\begin{document}
\ninept
\maketitle

\begin{abstract}
We introduce ESPnet-EZ, an extension of the open-source speech processing toolkit ESPnet, aimed at quick and easy development of speech models.
ESPnet-EZ focuses on two major aspects: (i) easy fine-tuning and inference of existing ESPnet models on various tasks and (ii) easy integration with popular deep neural network frameworks such as PyTorch-Lightning, Hugging Face transformers and datasets, and Lhotse.
By replacing ESPnet design choices inherited from Kaldi with a Python-only, Bash-free interface, we dramatically reduce the effort required to build, debug, and use a new model.
For example, to fine-tune a speech foundation model, ESPnet-EZ, compared to ESPnet, reduces the number of newly written code by 2.7x and the amount of dependent code by 6.7x while dramatically reducing the Bash script dependencies.
The codebase of ESPnet-EZ is publicly available.\footnote{\url{https://github.com/espnet/espnet}}
\end{abstract}



\section{Introduction} \label{sec:intro}
The \kaldi\ automatic speech recognition (ASR) toolkit \cite{kaldi} is one of the most successful open-source efforts in speech processing.
One of the key aspects of \kaldi's design comes from the \textit{recipe structure}.
Each recipe provides a fully reproducible experiment, such as handling data preparation, model training, and evaluation.
The various steps of the experiment often require the combination of various tools, including many Linux command-line executables and shell scripts/utilities. 
For example, \texttt{sox} is often used to handle various audio formats \cite{sox}, \texttt{awk} for data wrangling \cite{awk}, \texttt{make} for setting up development environments~\cite{make}, and NIST SCTK for ASR evaluation \cite{sctk}.
Bash scripts are used to glue these tools together. One prominent example is the \texttt{run.sh} ``main'' script inside each recipe directory which calls all the necessary tools and scripts and runs the recipe end-to-end.
This design enables easy reproduction and also encourages researchers to open-source their code, greatly benefiting the research community.
Additionally, this approach offers scalability and extensibility by piggybacking on the computational efficiency and wide availability of such tools.

The \espnet\ end-to-end speech processing toolkit \cite{watanabe2018espnet} aims to expand \kaldi's task coverage by supporting also other speech-related tasks \cite{hayashi2020espnet, hayashi2021espnet2, inaguma-etal-2020-espnet, li2020espnet, arora2021espnet, shi2022muskits, lu22c_interspeech, gao2023euro, jung2024espnet, yan2023espnet}. 
However, it fundamentally retains its core design principles. In fact, similarly to \kaldi, \espnet\ also focuses on end-to-end Bash script-based recipes to reproduce results from scratch.
For example, various challenges that cover a wide array of tasks are being held based on \espnet, such as multilingual automatic speech recognition (ASR)~\cite{shi2023findings}, robust speech recognition \cite{cornell2023chime}, and voice conversion \cite{zhao2020voice}.
Further, \espnet\ provides several recipes for training various speech foundation models from scratch.
These models are trained with large-scale data from various domains, requiring computational efficiency and parallelizability especially for data preparation and data loading. 
As such, these tasks are performed using many \kaldi\  tools, sacrificing ease of use for efficiency and training speed.
\espnet\ provides recipes for self-supervised speech models including HuBERT \cite{hsu2021hubert,chen23l_interspeech} and wav2vec-U \cite{baevski2021wav2vecu,gao2023euro}, weakly-supervised speech models such as Whisper \cite{radford2023robust} via OWSM~\cite{peng2023reproducing,peng2024owsm,tian2024owsm} and OWSM-CTC \cite{peng2024owsmctc}, as well as speech language models such as UniverSLU \cite{arora2023universlu} and VoxtLM \cite{maiti2024voxtlm}.

These foundation models are becoming the de-facto standard in various speech processing tasks, such as ASR, text-to-speech (TTS), or spoken language understanding (SLU).
They boast universal applicability on many tasks via fine-tuning, often achieving state-of-the-art performance, as demonstrated by benchmarks such as SUPERB \cite{yang2021superb}, ML-SUPERB \cite{shi2023ml, shi2024ml}, and SLUE \cite{shon2022slue}.
Thus, many increasingly rely on fine-tuning such large-scale models for new tasks rather than training from scratch.
Various methodologies can be used, such as full fine-tuning, parameter-efficient fine-tuning \cite{hu2021lora,he2024wav2gloss}, knowledge distillation \cite{choi2022temporal}, or prompting \cite{peng2023prompting}, making the approach less computationally demanding and more data-efficient.

\begin{figure}[t]
\vspace{-1em}
    \centering
    \subfloat[\# of new lines]{\includegraphics[height=0.2\textwidth]{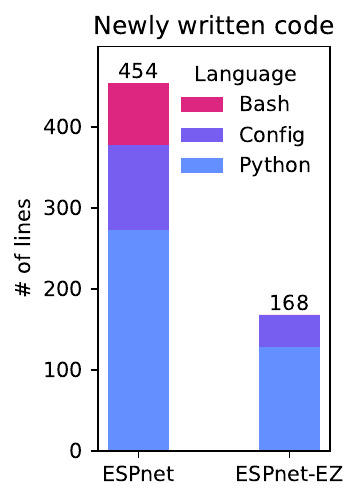}}
    \subfloat[\# of dependent files]{\includegraphics[height=0.2\textwidth]{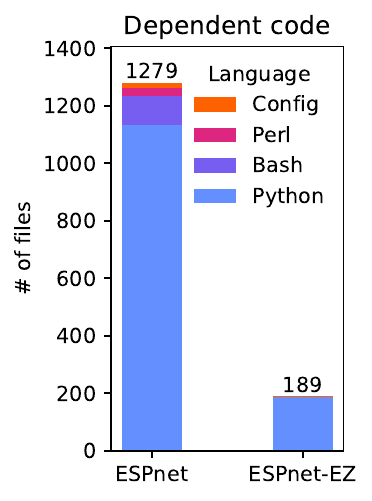}}
    \subfloat[\# of dependent lines]{\includegraphics[height=0.2\textwidth]{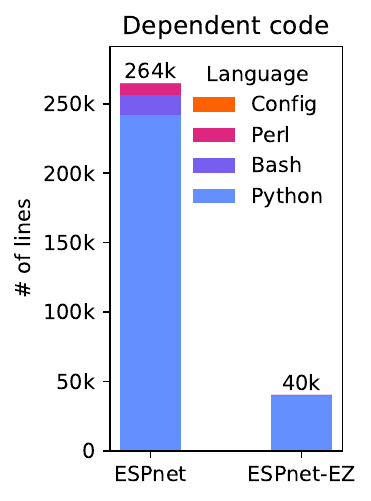}}
    \caption{
Quantitative comparison of \espnet\ and \espnetez.
We compare the case of fine-tuning the OWSM model, a speech foundation model, for the automatic speech recognition task on a custom dataset.
We use three criteria: (a) the number of new source code lines for the user to write and (b) the number of dependent source code files and (c) lines for each programming/scripting language.
We observe that \espnetez\ significantly reduces engineering efforts compared to the original \espnet.
Newly written lines are reduced by 2.7x, and the dependent code lines and number of files are reduced by 6.7x and 6.6x, respectively.
Further, \espnetez\ dramatically reduces the dependency on Bash and Perl.
}
\label{fig:code_lines}
\vspace{-1em}
\end{figure}
In addition, recent open-source projects focus on further reducing engineering complexity by leveraging Python as the go-to language, unlike \kaldi\ and \espnet, which still rely significantly on Bash and are thus arguably more difficult to debug.
Linux executables were the easiest way to integrate various tools into a single recipe.
However, many of these tools now have valid Python alternatives, e.g. \texttt{sox}  has \texttt{librosa} \cite{mcfee2015librosa} or \texttt{torchaudio} \cite{yang2022torchaudio}, \texttt{make} can be replaced by Python package managers such as \texttt{pip} \cite{pip} or \texttt{conda}~\cite{anaconda}, and NIST SCTK by \texttt{TorchMetrics} \cite{detlefsen2022torchmetrics} or \texttt{evaluate}~\cite{von2022evaluate}.
Further, replacing these with Python scripts facilitates debugging, readability, and integration with other Python-based frameworks. 
Other libraries that follow this same design philosophy are: \texttt{transformers} \cite{wolf2019huggingface}, \texttt{asteroid} \cite{Pariente2020Asteroid}, \texttt{NeMo} \cite{kuchaiev2019nemo}, \texttt{SpeechBrain} \cite{ravanelli2021speechbrain}, \texttt{WeNet} \cite{yao2021wenet}, and \texttt{S3PRL} \cite{yang2021superb}.

In summary, \espnet\ maintains a Bash script-based codebase for computational efficiency necessary for large-scale training.
However, it increases the engineering cost for use cases where efficiency is less critical, such as fine-tuning or integrating with other Python-based frameworks.
On the other hand, despite this drawback, \espnet\ has the most comprehensive coverage of speech tasks compared to all other Python-based frameworks.
These facts raise an interesting engineering question: \textit{Why not both? Can we inherit the tasks of \espnet\ while making it easier by replacing Bash with Python?}


To this end, we introduce \espnetez, an extension to \espnet, designed to reduce engineering efforts by removing the Kaldi-style recipe structure.
\espnetez\ inherits only the Python codebase of \espnet\ and provides a Python-only modular interface. 
While \espnet\  can focus on efficient training from scratch, \espnetez\ can provide ease of use via its Python-only codebase. 
As \espnet\ and \espnetez\ share the same Python codebase, \espnetez\ has native support for all \espnet downstream tasks: ASR, speaker verification, speaker diarization, speech enhancement, singing voice synthesis, text-to-speech, self-supervised learning, weakly supervised learning, language modeling, machine translation, speech-to-speech translation, speech-to-text translation, spoken language understanding, and unsupervised ASR.

Importantly, \espnetez\ removes all the \kaldi-style dependencies, avoiding shell-scripting altogether.
For example, if the user wants to fine-tune a speech foundation model, the user can install \espnet\  via a single line of code, \textit{i.e.} \texttt{pip install espnet}, avoiding a lengthy and convoluted build procedure.
Then, by importing the newly developed \espnetez\ library, the user can fine-tune the model with a Python-only trainer.
\espnetez\ removes the need to reformat the dataset into a \kaldi-style dataset and write dozens of new lines of bash scripts.
Moreover, we show it can be seamlessly integrated into existing machine learning Python frameworks, such as \texttt{PyTorch} \cite{paszke2019pytorch}, \texttt{PyTorch Lightning}, and \texttt{transformers Trainer} \cite{wolf2019huggingface} and supports also different dataset frameworks, such as vanilla \texttt{PyTorch} datasets, \texttt{Huggingface datasets} \cite{lhoest2021datasets}, and \texttt{Lhotse} \cite{zelasko2021lhotse}.

\begin{figure*}[th]
\vspace{-1em}
    \centering
    \includegraphics[width=0.98\textwidth]{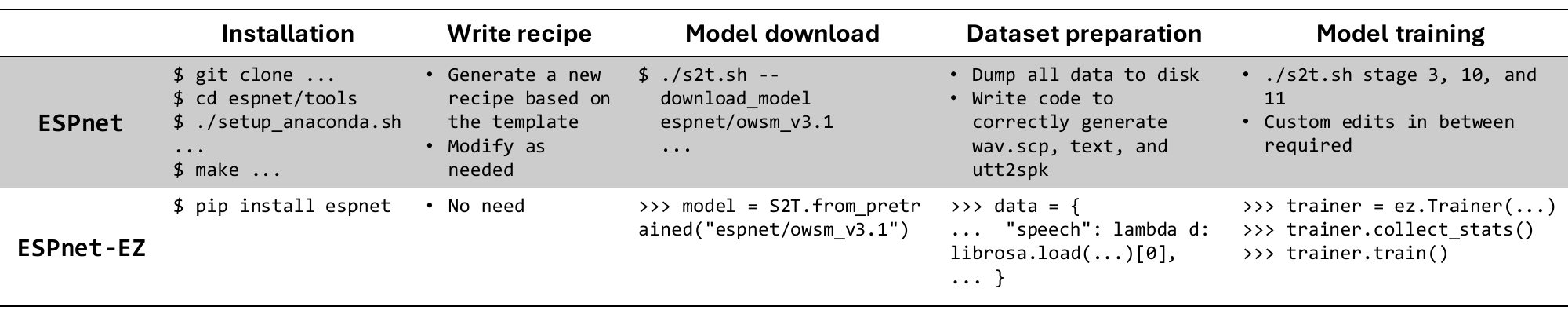}
    \caption{
Comparison of \espnet\ and \espnetez\ on fine-tuning the model with a custom dataset.
\espnet\ has to go through dozens of shell scripts and custom modifications, whereas all the codes in \espnetez\ are within a single Python script.
}
\label{fig:design_comp}
\vspace{-1em}
\end{figure*}

In the following, we first compare the original \espnet\ and \espnetez\ by taking a closer look at the use case of fine-tuning a speech foundation model (\Cref{sec:versus}).
Then, we conduct a comparative analysis between \espnet\ and \espnetez\ in \Cref{sec:comparison}.
Finally, we demonstrate various advantages of \espnetez, such as wide task coverage (\Cref{sec:demo}) and easy integration (\Cref{sec:integration}).
In summary, our work contributes the following: 
\begin{enumerate}
    \item Shows that \espnetez\  is far easier to use than \espnet\  via quantitative, qualitative, and user feedback comparisons.
    \item Demonstrates the wide task coverage of \espnetez.
    \item Provides various \espnetez\ examples on smooth integration with existing deep learning frameworks.
\end{enumerate}

\begin{figure}[th]
    \centering
    \includegraphics[width=0.48\textwidth]{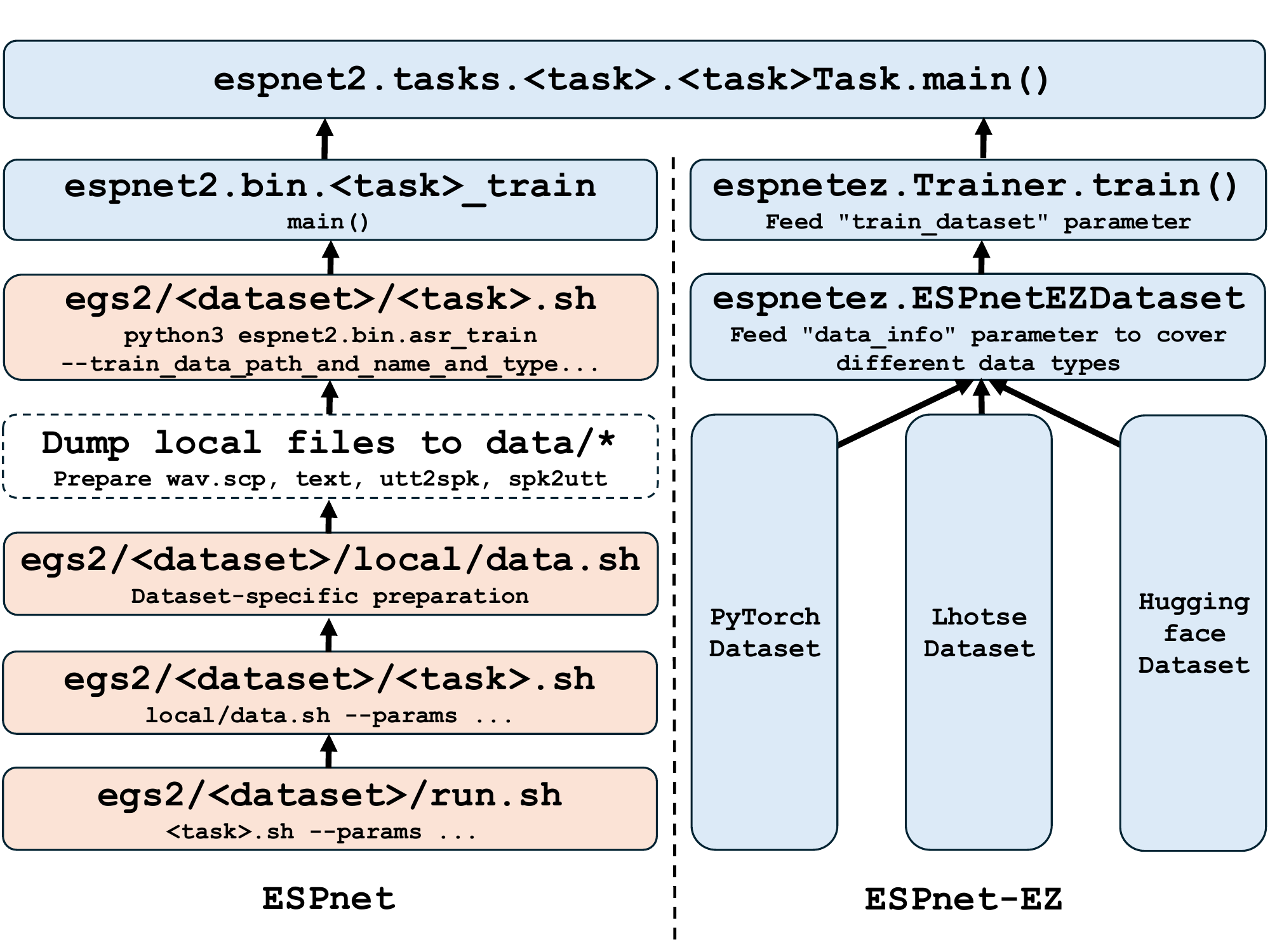}
    \caption{
Comparison of \espnet\ and \espnetez\ call stack of feeding the training data.
\espnet\ has to format the dataset into \kaldi-style and dump it to the local directories.
So, it introduces an implicit dependency between the previous data preparation step and the training step.
However, \espnetez\ avoids the stateful dependency via on-the-fly dataset preparation from Python-based data loaders.
}
\vspace{-1em}
\label{fig:design_comp_data}
\end{figure}

\section{Design of ESPNet-EZ}\label{sec:design}

\espnetez\ aims to make \espnet\ easier by removing  \kaldi-style dependencies and exposing \espnet\ core logic through a Python interface, thus allowing for fast development and easy integration with external toolkits. 
\Cref{fig:design_comp_data} compares implicit and explicit call stacks of \espnet\ and \espnetez.
\espnetez\ offers a more streamlined and shallower stack, inspired by deep learning frameworks like \texttt{PyTorch Lightning} or \texttt{transformers}.
It is built around two main modules: \texttt{Trainer} and \texttt{ESPNetEZDataset}, allowing flexibility through modularization.

\subsection{\texttt{Trainer}}
The core \texttt{espnet2.tasks}\ module is exposed to the user through a \texttt{Trainer} class interface. 
The interface is responsible for handling model training and fine-tuning for a particular task.
It covers the two \espnet\ recipe steps of collecting dataset statistics and training the model, where \espnet\ and \espnetez\ share the same core logic.
Currently, all 20 different speech tasks of \espnet\ are natively supported, but, differently from this latter,  it is also possible for the user to easily customize the \texttt{Trainer}\ for the application at hand. 



\subsection{\texttt{ESPNetEZDataset}}

\texttt{ESPNetEZDataset} is responsible for interfacing the data formatting for feeding into the training loop.
Original \espnet\ feeds the data formatting specifics via the command-line interface, limiting the extensibility and requiring the user to follow the \kaldi-style dataset preparation.
\texttt{ESPNetEZDataset} avoids the command-line interface via passing the Python function directly, removing the need of Bash script-based data processing steps altogether. 
\texttt{ESPNetEZDataset} is built on top of Pytorch dataset class, and thus is inhrently flexible in supporting also other dataset modules that are built in the same way. 
Prominent examples are Huggingface Datasets and Lhotse Datasets, which can be easily integrated with \espnetez. 
We provide more examples in \Cref{sec:integration}.



\section{Fine-tuning on ESPnet and ESPnet-EZ} \label{sec:versus}
Since a hands-on example is worth a million words, we outline the difference between \espnet\ and \espnetez\ regarding engineering efforts for a simple ASR fine-tuning pipeline.
The comparison is summarized in \Cref{fig:design_comp,fig:design_comp_data}.


\subsection{Fine-tuning on \espnet}\label{ssec:ft-espnet}
\textbf{Installation.}
For fine-tuning, the user has to create a new recipe in \espnet. 
However, \espnet\ has to be installed from source, which is a notoriously complex process due to its many dependencies. 
For example, for many recipes, one has to also download and build \kaldi\ beforehand.
\espnet\ provides a \texttt{Makefile} for installation.
However, depending on various Python environment types, such as system default Python, virtual environment, or \texttt{conda} environment, there are different installation procedures that need to be followed accordingly.
As these environments setups differ, builds often fail, requiring a lengthy debugging process through several build scripts.
Furthermore, many recipes require additional recipe-specific dependencies, which need to be manually installed by running additional scripts in \texttt{tools/installers}. 
As we can see, the installation procedure is quite complex, especially compared to recent frameworks such as \texttt{SpeechBrain} or \texttt{transformers}. 


\textbf{Writing a new recipe.}
The typical \espnet\ user leverages existing recipes to create a new fine-tuning recipe, such as OWSM and its \texttt{s2t1} pre-training recipe.
Nevertheless, it still requires substantial effort if the custom data or new task does not exactly match the pre-training recipe.  
As such, in \espnet\ the engineering effort for fine-tuning or training from scratch is largely the same.

\textbf{Model download.}
To download the model for fine-tuning, the user has to use a shell script with specific parameters.
For example, for OWSM, the user has to use the \texttt{s2t.sh} script inside the \texttt{s2st1} recipe.
Within this script, recipe-dependent file-linking happens, which increases the complexity of utilizing the model if the user wants to only perform inference in a standalone way.

\textbf{Dataset preparation.}
The dataset has to be prepared with \kaldi-compliant manifests, with specific filenames and rules to follow.
The actual rules are not implicitly written as a Python class but as a \kaldi\ reference document, making the dataset preparation more challenging.
Files such as \texttt{utt2spk} (mapping each utterance to the speaker) have to be taken care of, even if the user does not need additional functionalities (e.g. for blind speech separation or speech enhancement tasks). 
This further hinders usability when using outside dataset libraries.
For example, to leverage Huggingface's \texttt{datasets}, the user has to rewrite all the dataset rows into separate audio files.
This means that another copy of the dataset in the disk is required, thus using extra storage.
Furthermore, additional data preparation steps after such manifest preparation are usually required.
For example, in the case of OWSM's \texttt{s2t1} recipe, the dataset has to go through additional formatting (Stage 3), filtering (Stage 4), and needs custom code for tokenization of the transcript.
To go through these stages, the user has to feed dozens of hyperparameters to the \texttt{s2t.sh} script.

\textbf{Model training and inference.}
For various batching strategies, \espnet\ has to collect training data statistics (Stage 10).
As the existing recipe is not designed for fine-tuning, the user additionally has to handle also this statistics collection step.
Only after this latter step, based on the configuration file, the model can be fine-tuned.
However, once the fine-tuning is over \espnet\ actually lacks a straightforward way to run inference due to the data preparation steps needed. To prepare test data, the user has to go through the same lengthy procedure as done for the training data.

\subsection{Fine-tuning on \espnetez} \label{ssec:ft-espnetez}
\textbf{Installation.}
To install and use \espnetez, the user only needs a Python package manager, regardless of the environment.
After activating a Python environment of any liking, typing \texttt{pip install espnet} will finish the installation process.

\textbf{Recipe preparation.}
\espnetez\ does not depend on \kaldi-style recipes.
Rather, it gives the user maximum freedom on how to use the \espnet\  library for their own projects.
In later sections, we provide easy ways to use the model as-is in the existing training code or train the model based on the \espnetez\ trainer functionality.

\textbf{Model download.}
\espnetez\ removes the Bash scripts surrounding the \texttt{ESPnet-model-zoo}, which is previously introduced in \espnet\ for easier model download.
Hence, users can easily download and use the models via \texttt{from\_pretrained} method within the Python codebase, similar to \texttt{asteroid} and \texttt{transformers}.
Model download and management are performed under the hood, reducing engineering overhead.

\textbf{Dataset preparation.}
As \kaldi-style dataset preparation is not anymore necessary (but still supported by \espnetez), various dataset frameworks can be easily integrated.
As the \espnet\ model receives PyTorch tensors, the user can easily build a shallow connector for existing dataset frameworks, such as vanilla PyTorch datasets, Huggingface \texttt{datasets}, and Lhotse, to feed the data into the model.

\textbf{Model training and inference.}
Depending on the use case, the user can choose between leveraging the \espnetez\  \texttt{Trainer} or their custom trainer for training and inferencing the model.
The \espnet\  model is based on \texttt{PyTorch}, so it can be seamlessly integrated into existing frameworks, such as \texttt{PyTorch Lightning}.
\espnet\  already features a Python-friendly implementation of inference; therefore, we can directly incorporate the inference code with \espnetez\  scripts.

\section{Comparing ESPnet and ESPnet-EZ} \begin{figure}[t]
\vspace{-1em}
    \centering
    \subfloat[\espnetez\ over \espnet]{\includegraphics[height=0.35\columnwidth]{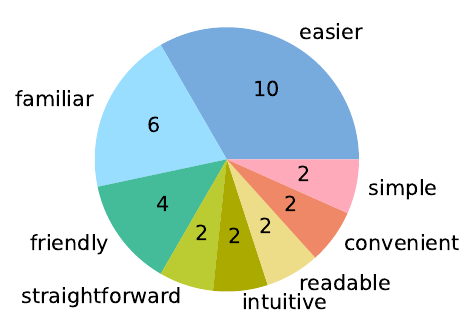}}
    \subfloat[\espnet\ over \espnetez]{\includegraphics[height=0.35\columnwidth]{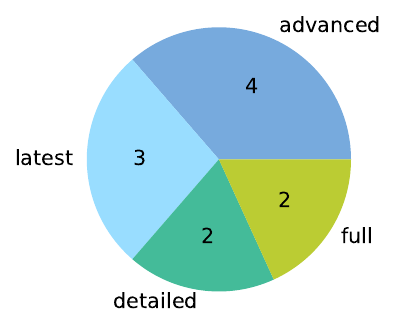}}
    \caption{Summary of user feedbacks on the benefits of using \espnetez\ or \espnet.}
    \label{fig:feedback}
\end{figure}

\label{sec:comparison}

\subsection{Qualitative Comparison (High-level Summary)}
As said, the current \espnet\  codebase is optimized to be highly efficient in cluster environments by leveraging various shell scripts, making it well suited for large-scale training.
However, as said, leveraging these foundation models in an off-the-shelf manner is equally important as it is arguably the mainstream approach for solving various speech tasks.
Hence, \espnetez\ reduces the friction of leveraging existing models by reimplementing some of the shell-based codebases in a Python-only manner.
Even though it introduces unavoidable computational overhead, it is negligible for smaller-scale training and fine-tuning (e.g. for Librispeech \cite{panayotov2015libri}).

\subsection{Quantitative Comparison}
We further quantify the differences by measuring engineering efforts in \Cref{ssec:ft-espnet,ssec:ft-espnetez}.
As an indirect metric, we measure the amount of code in various ways.
We count the number of dependent files within the \espnet\ codebase, the number of lines of the files, and the newly written number of lines.
For the \texttt{espnet}, \texttt{espnet2}, and \texttt{espnetez} packages, we traverse through all the imported dependencies.
For the command line tools for \espnet, we count all codes in the \texttt{script} and \texttt{pyscript} directory within the recipe.
We summarize the results in \Cref{fig:code_lines}.
We can immediately observe that the amount of code is dramatically reduced.
Furthermore, the language distribution becomes more Python-friendly, effectively removing Bash and Perl script dependencies altogether.

\subsection{User Feedback Comparison}
We gathered real-world feedback from the \espnet\ users.
We provided a \texttt{Jupyter} notebook demonstration similar to \Cref{subsec:slu} and asked the users to list the pros and cons of using \espnetez relative to the full \espnet\ toolkit.
There were 21 respondents, encompassing undergraduate students who are new at speech research and graduate students who have already published multiple papers.
Each respondent already had basic experience on \espnet, where they went through the notebook demonstration of the \espnet\ ASR recipe.
We summarize the responses in \Cref{fig:feedback} and include all raw responses in the supplementary material.
To obtain the summaries, we use a pre-trained Part-Of-Speech Tagger\footnote{\url{https://github.com/explosion/spaCy}} to parse out adjectives. 
After grouping similar words, we counted each word that occurred more than once.
The feedback to \espnetez\ was generally positive, as the respondents found that its similarities with \texttt{transformers}, make it easier and familiar to use. 
However, some of the students also acknowledged that different features and customization may be relatively limited in \espnetez, and that fully leveraging the capabilities of \espnetez\  may still require deep understanding of the \espnet\ package.
The responses align well with our design decision: \espnetez\ handles the easier use cases, while \espnet\ covers more challenging scenarios.

\section{Task Coverage of ESPnet-EZ} \label{sec:demo}
In this section, we demonstrate the wide applicability of \espnetez\  to various tasks. 
In detail, we fine-tune the OWSM speech foundation model in various cases, such as (i) both the dataset and the task are seen during training (\Cref{subsec:asr,subsec:st}), (ii) the dataset is seen but the task is unseen (\Cref{subsec:tts}), (iii) the dataset is unseen but the task is seen (\Cref{subsec:w2g}), and (iv) both the dataset and the task are unseen (\Cref{subsec:slu,subsec:w2g}).
Specifically, we fine-tune the \texttt{v3.1-ebf-base} model\footnote{\url{https://huggingface.co/espnet/owsm_v3.1_ebf_base}} \cite{peng2024owsm} with 101M parameters.
We also include the reported performance of state-of-the-art models for comparison on each task.
All experiments are implemented via \espnetez, where we provide all the tutorials and codebase.\footnote{\url{https://espnet.github.io/espnet/notebook/ESPnetEZ/}}




\subsection{Automatic Speech Recognition (ASR)}\label{subsec:asr}
\begin{table}[ht]\centering
\scriptsize
\vspace{-1em}
\begin{tabular}{lrrrrrr}\toprule
&\multicolumn{2}{c}{dev (WER $\downarrow$)} & &\multicolumn{2}{c}{test (WER $\downarrow$)} \\\cmidrule{2-3}\cmidrule{5-6}
&clean &other & &clean &other \\\midrule
WavLM-Base+ \cite{chen2022wavlm} & N/A & N/A & & 4.6 & 10.1 \\
WavLM-Large (with LM) \cite{chen2022wavlm} & N/A & N/A & & 2.1 & 4.0 \\ \midrule
Baseline  &4.1 &9.1 & &4.0 &9.9 \\
FT with \espnet  & 4.5 & 9.8 & & 4.8 & 10.3 \\
FT with \espnetez   &3.4 &8.5 & &3.5 &9.0 \\
FT with \espnetez + \texttt{Lhotse}  & 3.5 & 8.8 & & 3.7 & 9.5 \\
LoRA with \espnetez  &3.4 &8.4 & &3.6 &8.8 \\
\bottomrule
\end{tabular}
\caption{
WER on Librispeech dev/test sets.
\textbf{Baseline} represents the original model before fine-tuning.
\textbf{FT} and \textbf{LoRA} indicates full-finetuning and PEFT results.
\texttt{Lhotse} denotes online augmentation through \texttt{Lhotse} \texttt{dataio} functions. 
}
\label{tab:asr_result}
\end{table}
\textbf{Task definition.}
The model has to generate a transcription from a given speech input.
We fine-tuned the model using a 100h train-clean-100 training subset of LibriSpeech \cite{panayotov2015libri}.
For early stopping, we use dev-clean and dev-other subsets.

\textbf{Baselines.}
We compare the performance of the OWSM model before fine-tuning, fine-tuned model by \espnet\ or \espnetez.
For the ground truth of \espnet, we use the original input format with timestamps per the existing tutorials, while \espnetez\ uses the format with transcription only.
For \espnetez, we additionally employ \texttt{Lhotse} for online data augmentation and Low-Rank Adaptation (LoRA) \cite{hu2021lora} for parameter-efficient fine-tuning (PEFT).
Finally, we compare with the two state-of-the-art results from WavLM \cite{chen2022wavlm}.
WavLM-Base+ (95M) has a similar parameter size as OWSM, and WavLM-Large (316M), with 3x the parameter size, uses shallow fusion with the transformer language model (LM).

\textbf{Fine-tuning details.}
We conduct a grid search to find the optimal hyperparameter.
We use the AdamW optimizer \cite{adamw} with the learning rate grid of [2e-3, 1e-3, 5e-4, 1e-4] and the warm-up step grid of [5000, 15000].
The optimal configuration was the learning rate of 1e-4 and 15000 warm-up steps.
The model was trained for 10 epochs.
We use the same hyperparameter for both \espnet\ and \espnetez.
For LoRA, we use a rank of 8 and an alpha of 8.
For augmentation, we use the speed, volume, and tempo augmentations for 30\% of the data during training.
Volume can be either 10\% increased or decreased, and others can be 10\% decreased.

\textbf{Results.}
\Cref{tab:asr_result} demonstrates improvements in Word Error Rate (WER) of the fine-tuned model.
Even though LibriSpeech is used during OWSM training, the transcription format is different.
Hence, the model is likely to be fine-tuned towards the new format.
However, \espnet\ fine-tuning decreases performance, partially because the obtained hyperparameters being suboptimal in transcriptions that include timestamps.
Also, LoRA is comparable to full fine-tuning, demonstrating the effectiveness of PEFT.
However, online augmentation of \texttt{Lhotse} slightly degraded the performance, possibly due to Librispeech being relatively clean.
Finally, fine-tuning OWSM shows superior performance over WavLM-Base+, which has a similar setting.
The performance difference with WavLM-Large implies that scaling up to bigger OWSM models and shallow fusion with LM may further improve performance.

\subsection{Speech Translation (ST)}\label{subsec:st}

\begin{table}[!htp]\centering
\vspace{-1em}
\scriptsize
\begin{tabular}{lrrrr}\toprule
&BLEU $\uparrow$ &chrF $\uparrow$ &TER $\downarrow$\\\midrule
Yan et al. \cite{yan-etal-2023-ctc} & 29.2 & N/A & N/A \\\midrule
Baseline &21.4 &49.0 &71.2 \\
FT &23.3 &49.7 &57.5 \\
LoRA &20.2 &45.7 &59.6 \\
Cascaded & 18.5 & 46.0 & 73.5 \\
\bottomrule
\end{tabular}
\caption{
Performance on MuST-C v2 English-to-German test set.
\textbf{Baseline} represents the original model before fine-tuning.
\textbf{FT} and \textbf{LoRA} indicates full-finetuning and PEFT results.
For \textbf{Cascaded}, \texttt{whisper-tiny} and \texttt{t5-base} model is used.
}
\label{tab:st_table}
\end{table}
\textbf{Task definition.}
The model has to generate a translated text from speech from one language to another.
OWSM model directly translates the audio input into the target language without generating the source language transcription.
To evaluate fine-tuning performance on English-to-German translation, we used the MuST-C-V2 \cite{cattoni2021mustc} dataset, which includes TED talks and their transcriptions.
For evaluation, we use BiLingual Evaluation Understudy (BLEU)~\cite{papineni2002bleu}, CHaracter-level F-score (chrF)~\cite{popovic2015chrf}, and Translation Error Rate (TER)~\cite{snover2006study}.
We use SacreBLEU~\cite{post2018call} for valid set.

\textbf{Baselines.}
Similar to \Cref{subsec:asr}, we compare the performance of OWSM before and after fine-tuning.
We test full fine-tuning and LoRA.
Also, we train a cascaded ST model, employing \texttt{whisper-tiny} \cite{radford2023robust} (transcribes speech into English) and the \texttt{t5-base} \cite{raffel2020exploring} (translates English to German) from \texttt{transformers}.

\textbf{Fine-tuning details.}
We go through the same hyperparameter search process of \Cref{subsec:asr}.
For LoRA, we use learning rate of 2e-3.
The prompt template for \texttt{whisper-tiny} is \texttt{translate English to German: <ASR result>}, obtained from the official website.\footnote{\url{https://huggingface.co/openai/whisper-tiny}}
We also format the first character in uppercase and the rest in lowercase.
We compared with the state-of-the-art model \cite{yan-etal-2023-ctc} with ST-specific architecture design and 72M parameters.

\textbf{Results.}
\Cref{tab:st_table} shows that fine-tuning improves BLEU, while LoRA shows slight degradation.
Nevertheless, LoRA improves TER by 11.6 points, indicating over the baseline.
Cascaded ST performance was suboptimal, even though the total number of parameters (261M) is larger than OWSM.
However, fine-tuned OWSM model was suboptimal relative to the state-of-the-art, where the encoder module of Yan et al. \cite{yan-etal-2023-ctc} is engineered to handle both source and target language features independently.
This architectural distinction, absent in OWSM, facilitates the acquisition of target language information, potentially contributing to enhanced translation.

\subsection{Spoken Language Understanding (SLU)} \label{subsec:slu}
\begin{table}[!htp]\centering
\vspace{-1em}
\scriptsize
\begin{tabular}{lcc}\toprule
&Accuracy $\uparrow$\\\midrule
UniverSLU \cite{arora2023universlu} & 90.3\% \\
\midrule
FT  &82.45\% \\
LoRA  & 65.89\% \\
\bottomrule
\end{tabular}
\caption{
Accuracy of the intent classification task on SLURP dataset.
\textbf{FT} and \textbf{LoRA} indicates full-finetuning and PEFT results.
}
\label{tab:slu_result}
\end{table}
\textbf{Task definition.}
The model has to extract the meaning given the spoken content.
For the SLU task, we evaluated the model's performance using the SLURP \cite{bastianelli2020slurp} dataset, consisting of various in-home prompts for home assistants.
We especially focus on the intent classification task, which is not included during OWSM training.

\textbf{Baselines.}
As OWSM does not support the SLU task, we only test the two variants of fine-tuning: full fine-tuning and LoRA.
We compare the results with UniverSLU \cite{arora2023universlu}, the state-of-the-art model that fine-tunes Whisper-medium (769M) \cite{radford2023robust} to 12 SLU tasks.

\textbf{Fine-tuning details.}
We go through the same hyperparameter search of \Cref{subsec:asr} and set the learning rate to 5e-4.
For warm-up, we choose 5k and 15k steps for full-finetuning and LoRA.
We add a special task token \texttt{<intent>} to adapt the model to the new task.

\textbf{Results.}
\Cref{tab:st_table} shows that OWSM can be adapted to the unseen task with fine-tuning, even though it requires full fine-tuning for better performance.
We observed that OWSM outputs several tokens to express intent, introducing errors while decoding the intent labels.
Also, the model fitted to the training data too quickly.
The empirical observations further support the multi-task approach of UniverSLU to increase the training data size.

\subsection{Text to Speech (TTS)}\label{subsec:tts}
\begin{table}[!htp]\centering
\scriptsize

\vspace{-1em}
\begin{tabular}{lccccc}\toprule
 & MCD $\downarrow$ & F\(_0\) Corr $\uparrow$ & SECS $\uparrow$ & UTMOS $\uparrow$ \\\midrule
USAT \cite{10508477} & N/A & N/A & N/A & 3.81 \\\midrule
Baseline  & 11.54 & 0.1499 & 0.0677 & 3.13  \\
FT  & 7.90 & 0.2744 & 0.5494 & 3.44  \\
\bottomrule
\end{tabular}


\caption{
Performance on VCTK test set.
\textbf{Baseline} and \textbf{FT} represents the performance before and after full fine-tuning.
}
\label{tab:tts_result}
\end{table}
\begin{table*}[ht]\centering
\scriptsize
\begin{tabular}{lccccccccccc}\toprule
           & \multicolumn{2}{c}{Transcription (CER $\downarrow$)} && \multicolumn{2}{c}{Underlying (CER $\downarrow$)} && \multicolumn{2}{c}{Gloss (CER $\downarrow$)} && \multicolumn{2}{c}{Translation (chrF++ $\uparrow$)} \\ \cmidrule{2-3}\cmidrule{5-6}\cmidrule{8-9} \cmidrule{11-12}
           & Seen           & Unseen           && Seen          & Unseen         && Seen       & Unseen       && Seen          & Unseen          \\
           \midrule
\texttt{v3.1-ebf-base} \cite{peng2024owsm,he2024wav2gloss}  & 48.2 & 67.7 && 54.8 & 80.0 && \textbf{75.0} & \textbf{102.9} && \textbf{13.7} & \textbf{11.6} \\
\texttt{v3.1-small} \cite{peng2024owsm}  & 48.2 & 67.3 && \textbf{47.8} & \textbf{75.0} && 76.4 & 113.1 && 13.6 & 10.4 \\
\texttt{v3.2-small} \cite{tian2024owsm} & \textbf{43.1} & \textbf{67.1} && 48.5 & 76.4 && 83.0 & 121.7 && 13.1 & 10.1 \\
\bottomrule
\end{tabular}
\caption{
Performance on four subtasks within the Fieldwork corpus.
OWSM variants are fine-tuned for each of the subtasks.
We import the results from \cite{he2024wav2gloss} for the \texttt{v3.1-ebf-base} model.
Metrics are averaged per language.
Unseen indicates languages unseen during training.
}
\vspace{-1em}
\label{tab:w2g_result}
\end{table*}

\textbf{Task definition.}
The task aims to synthesize speech from text in the voice of the target speaker.
Following \cite{hayashi2020espnet}, we evaluate performance using Mel Cepstral Distortion (MCD) \cite{407206}, F\(_0\) Pearson correlation coefficient (F\(_0\) Corr), speaker embedding cosine
similarity (SECS), and UTMOS~\cite{saeki2022utmos}.
For SECS, we use the speaker embeddings of ESPnet-SPK model \cite{jung2024espnet} trained on VoxCeleb \cite{nagrani2017voxceleb, chung2018voxceleb2}.


\textbf{Baselines.} We use a VITS model pre-trained on LibriTTS \cite{zen19_interspeech} as our baseline and fully fine-tune on VCTK \cite{veaux2017cstr}.
We compare our results with the state-of-the-art USAT model \cite{10508477}.

\textbf{Fine-tuning details.}
For the task, we focus on fine-tuning the model, where a source TTS model is adapted to an unseen dataset containing unknown speakers.
We use the same hyperparameters as pre-training, except for the AdamW optimizer with a learning rate of 2e-4 and the batch size of 1 due to GPU memory constraints.
Following the existing ESPnet-TTS recipe~\cite{hayashi2020espnet}, we use 540 samples of VCTK for dev set, 540 samples for test set, and the rest for training. 

\textbf{Results.}
\Cref{tab:tts_result} demonstrates improved audio quality across all metrics after adaptation via fine-tuning.
While the fine-tuned model does not achieve state-of-the-art performance, the results remain promising particularly given the constrained conditions, i.e., with a vanilla VITS architecture and limited data sources.






\subsection{Low-resource Languages}\label{subsec:w2g}
\textbf{Task definition.}
The Fieldwork corpus \cite{he2024wav2gloss} provides linguistically rich annotations of 37 endangered languages.
It contains four subtasks: transcription, underlying, gloss, and translation.
Underlying and gloss can be understood as an intermediate task between transcription and translation.
For the transcription task (ASR), the model has to use the orthography of each language.
The underlying task is similar to the transcription task, where the model has to additionally segment by morpheme with the underlying representation before applying the phonological rules.
The interlinear gloss task has the same segment as the underlying task, where the gloss additionally requires brief explanations for each morpheme. 
Target language of the translation task (ST) is English, which is evaluated by the modified character-level F-score (chrF++) \cite{popovic2017chrf}.
All other tasks are evaluated by the character error rate (CER).

\textbf{Fine-tuning details.}
\cite{he2024wav2gloss} provides baselines using the OWSM \texttt{v3.1-ebf-base} model, fine-tuning the model on each of the tasks.\footnote{\url{https://github.com/juice500ml/finetune_owsm}}
For our experiments, we test bigger variants of OWSM, namely, \texttt{v3.1-small} \cite{peng2024owsm} and \texttt{v3.2-small} \cite{tian2024owsm}.
\texttt{v3.2-small} differs with \texttt{v3.1-small} by the training data preprocessing strategy.
For both \texttt{v3.1-small} and \texttt{v3.2-small} with larger parameter size (367M), we use a smaller learning rate of 1e-4 and batch size of 8 to meet the GPU constraints and stabilize the training.
To make the number of iterations the same, we use 5 epochs.

\textbf{Results.}
In \Cref{tab:w2g_result}, \texttt{v3.2-small} shows great performance in transcription, likely due to careful curation of the training data.
However, the smallest \texttt{v3.1-ebf-base} performs best in both gloss and translation, arguably a more challenging task than transcription.
One possible explanation is the limited number of epochs, where bigger models potentially require more iterations to train.

\section{Easy Integration of ESPnet-EZ} \label{sec:integration}
In this section, we go over various examples where \espnetez\ integrated with existing deep learning and dataset frameworks.

\subsection{End-to-end Training with Multiple Models}\label{subsec:cascade}
One of the advantages of a Python-only codebase is that it allows to easy integration of pre-trained models from different frameworks.
In \Cref{subsec:st}, we demonstrate the case where the cascaded models' components are from two different frameworks:  \espnet\ and \texttt{transformers}.
As both models are PyTorch-based, we can train both with the \espnetez\ \texttt{Trainer}. This demonstrates  the flexibility of \espnetez\ Python-only codebase.

\subsection{Applying Audio Augmentations}\label{subsec:augmentation}
As \espnetez's dataset is a shallow wrapper for \texttt{PyTorch}'s vanilla datasets, one can easily apply various techniques during training.
The user can choose either the dataset or the data loader (which includes batching techniques) to be inputted in the \espnetez\ trainer.
For example, we demonstrate on-the-fly speech augmentation in \Cref{subsec:asr}.
We apply various augmentation techniques from \texttt{Lhotse} while using the \espnetez\ trainer.

\subsection{Integration with External Frameworks}\label{subsec:lightning}
In \Cref{fig:design_comp_data}, we observe that we can also skip the \espnetez\ \texttt{Trainer} when training the model.
It is demonstrated in \Cref{subsec:w2g} where we swap \espnetez\ \texttt{Trainer} with the trainer from \texttt{PyTorch-Lightning}.

\subsection{Online Fine-tuning Demo}
We implemented a fine-tuning demonstration on Huggingface Spaces.\footnote{\url{https://huggingface.co/spaces/ms180/owsm_finetune}}
Fine-tuning can be performed simply by uploading training data in a ZIP archive.
The demo showcases the great integration ability of \espnetez, allowing users to evaluate the performance of the fine-tuned OWSM on a custom dataset within a few minutes.
\section{Conclusion}\label{sec:concl}
In this work, we introduce \espnetez, an extension of the open-source speech processing toolkit \espnet.
\espnetez\ focuses on easier fine-tuning, inference, and integration with other deep learning frameworks.
By removing the \kaldi-style dependencies,  \espnetez\ provides a Python-only \espnet\ interface, simplifying the process of implementation while maintaining all \espnet\ task coverage.
We provide various comparative analyses, demos, and experiments to corroborate the advantages above.

\section{Acknowledgements}\label{sec:ack}
This work used the Bridges2 system at PSC and Delta system at NCSA through allocation CIS210014 from the Advanced Cyberinfrastructure Coordination Ecosystem: Services \& Support (ACCESS) program, which is supported by National Science Foundation grants \#2138259, \#2138286, \#2138307, \#2137603, and \#2138296.

\bibliographystyle{IEEEbib}
\bibliography{strings,refs}

\newpage
\section{Supplementary Materials}
\subsection{User feedbacks}\label{ssec:feedback}
\subsubsection{Pros}
\begin{itemize}
    \item It was much easier to write, understand and update pythonic code according to my requirements instead of bash scripts
    \item The compatability with Hugging Face made it far less initmidating, as I could transfer my knowledge of the Hugging Face datasets library and trainer () API to ESPnet. I imagine this would make it a much easier entry point for beginners as well.
    \item It is user friendly because everything can be done with Python. Unlike previous demos where we had to edit bash files, we were able to easily change the parameters and fine-tune the model with python scripts
    \item ESPnet EZ is quite easy to use, without the need to actually touch the low-level details of ESPnet.
    \item Espnet EZ is a more straightforward tool to work with.
    \item Staying within the python to make config changes makes it very simple to implement configuration changes.
    \item My experience with ESPnet EZ is very good
    \item ESPnet EZ provides a more convient way to use ESPnet. It is a bit similar to Hugging Face abstractions
    \item ESPnet EZ is easier to use and has less complexity. It has a simpler configuration and easier setup, making it more acesssible to beginners or those interested in fast implementation
    \item ESPnet-EZ comes with predefined pipelines and pretrain model loading, making it easier to perform common speech processing tasks without needing to train models from scratch
    \item I love it. It is much more similar to other tools I have use before. I instantly recognize the Trainer class as similar to the class of the same name from the transformers library for text.
    \item I think it is easier for implementing
    \item Espnet EZ significantly lowers the entry barrier for useres, especially those familiar with python and huggingface frameworks as it is less script heavy
    \item It has more intuitive usage of popular frameworks, making it easier for newcomers to grasp and utilize
    \item The codebase of ESPnet EZ is more readable and interpretable compared to the traditional ESPnet toolkit
    \item Its alignment with mainstream deep learning practices, where users can easily understand the structure and flow of the code, facilitating easier customization and troubleshooting
    \item Since everything was python based it made it much easier to use.
    \item It helps a lot on understanding how to use ESPnet
    \item It allows user to implement speech processing capabilities quickly and with minimal setup
    \item Has a more user-friendly iterface, simplified workflow for common tasks
    \item Convenient toolkit for applying some finetuning techniques like LORA
    \item ESPnet EZ is really user friendly
    \item Dataset format conversion is straightforward
    \item Training configuration is readable
    \item It feels more intuitive because it is python based
    \item EZ simplifies ASR tasks with user-friendly iteraces and pre-trained models
\end{itemize}

\subsubsection{Cons}
\begin{itemize}
    \item Lack of evaluation framework is the most apparent issue. Having to rely on accuracy plots each time you change hyperparameters is cumbersome
    \item Additionally, there is no support for hyperparameter optimization, which can be reaslly useful for ablation studies for research projects
    \item Advanced users who require more ocntrol over various parameters and models may find it limited in terms of customization. Due to its simplication, it may not support all the advanced features and functionalilites
    \item It lacks fine-grained control of the entire training procedure. When we want to add a new pipelien that is not in ESPnet,, we still need to go back to the original ESPnet.
    \item Using the bash scripts might have the advantage of providing more control over how the different files are run
    \item It may limit the flexibility and customizability available to users. For some detaile and complex settings, I would choose ESPnet
    \item ESPnet EZ might not keep up with the latest updates as quickly as the main ESPnet toolkit
    \item It might be harder to debug if any unsuspected things happen.
    \item I would imagine ESPnet is more robust and potentially slightly faster to compensate for its difficulty
    \item It comes as the cost of the depth of customization and the range of features
    \item Less flexibility for advanced users and possibly limited options for customizations
    \item Not sure if EZ supports building a new module based on an existing model with some small changes (such as adding a linear layer at the end)
    \item Same as python versus shell
    \item May offer limited customization and control compared to full ESPnet
\end{itemize}

\end{document}